\documentclass{PoS}

\usepackage{amsmath}
\usepackage{slashbox}

\PoS{PoS(HEP2005)063}

\title{Unifying the Fixed Order Evolution of Fragmentation Functions 
with the Modified Leading Logarithm Approximation}

\ShortTitle{Unifying Fixed Order Evolution with the MLLA}

\author{\speaker{Simon Albino}\\
        2nd Institute for Theoretical Physics, Hamburg University, Germany\\
        E-mail: \email{simon@mail.desy.de}}

\author{Bernd Andreas Kniehl\\
        2nd Institute for Theoretical Physics, Hamburg University, Germany}

\author{Gustav Kramer\\
        2nd Institute for Theoretical Physics, Hamburg University, Germany}

\author{Wolfgang Ochs\\
        MPI f\"ur Physik, Munich, Germany}

\abstract{An approach which unifies
the Double Logarithmic Approximation at small $x$ and the leading order DGLAP evolution of
fragmentation functions at large $x$ is presented. This approach reproduces
exactly the Modified Leading Logarithm Approximation, but is more complete due to the
degrees of freedom given to the quark sector and the inclusion
of the fixed order terms. We find that data from the largest $x$ values
to the peak region can be better fitted than with other approaches.}

\FullConference{International Europhysics Conference on High Energy Physics\\
		 July 21st - 27th 2005\\
		 Lisboa, Portugal}

\begin{document}

The current optimum description of single hadron inclusive production is provided
by the QCD parton model, which requires 
fragmentation functions (FFs) $D_a^h(x,Q^2)$ describing the probability for a parton
$a$ to emit a hadron $h$ carrying a fraction $x$ of its momentum.
Omitting the $h$ and $a$ labels,
the evolution of the FFs in the factorization scale $Q^2$ 
at large and intermediate $x$ is well described \cite{KKP2000} by the leading order (LO) 
DGLAP equation \cite{DGLAP}
\begin{equation}
\frac{d}{d\ln Q^2} D(x,Q^2)=\int_x^1 \frac{dz}{z}a_s(Q^2) P^{(0)}(z) D\left(\frac{x}{z},Q^2\right),
\label{DGLAPx}
\end{equation}
where $P^{(0)}(z)$ are the LO splitting functions
calculated from fixed order (FO) pQCD, and $a_s(Q^2)=\alpha_s(Q^2)/(2\pi)$.
As $z\rightarrow 0$, the LO splitting function $a_s P^{(0)}(z)$ diverges due to terms
of the form $a_s/z$. These double logarithms (DLs) occur at all orders in the FO
splitting function, being generally of the form $(1/z) (a_s \ln z)^2 (a_s \ln^2 z)^r$ for
$r=-1,...,\infty$.
As $x$ decreases, Eq.\ (\ref{DGLAPx}) will therefore become a poor approximation
once $\ln (1/x) = O(a_s^{-1/2})$. A small $x$ description is obtained by resumming these DLs 
using the Double Logarithmic Approximation
(DLA) \cite{Bassetto:1982ma;Fadin:1983aw}, given by
\begin{equation}
\frac{d}{d \ln Q^2}D(x,Q^2)=\int_x^1 \frac{dz}{z} \frac{2C_A}{z}A  z^{2\frac{d}{d\ln Q^2}}
\left[a_s(Q^2) D\left(\frac{x}{z},Q^2\right)\right], \qquad
A=\left( \begin{array}{cc}
0 & \frac{2 C_F}{C_A} \\
0 & 1
\end{array} \right)
\label{DLAx}
\end{equation}
for $D=(D_{\Sigma},D_g)$, where $D_{\Sigma}=\frac{1}{n_f}\sum_{q=1}^{n_f} (D_q 
+D_{\overline{q}})$ is the singlet FF.
The evolution of the valence quark and non-singlet FFs vanishes in the DLA.

We now construct an approach suitable for large and small $x$ simply and consistently 
by using the DLA to resum the DLs in the DGLAP evolution, 
as described in more detail in Ref.\ \cite{Albino:2005gg}.
We will use Eq.\ (\ref{DLAx}) to modify $a_s P^{(0)}$ in Eq.\ (\ref{DGLAPx}) to
\begin{equation}
a_s P^{(0)}(z) \rightarrow P^{\rm DL}(z,a_s)
+a_s \overline{P}^{(0)}(z),
\label{replace}
\end{equation}
where $P^{\rm DL}(z,a_s)$ contains the complete DL contribution, 
while $a_s\overline{P}^{(0)}(z)$ is obtained
by subtracting the LO DLs, already accounted for in $P^{\rm DL}$,
from $a_s P^{(0)}(z)$ to prevent double counting.
To obtain $P^{\rm DL}$, we work in Mellin space, 
$f(\omega)=\int_0^1 dx x^{\omega} f(x)$.
Upon Mellin transformation, Eq.\ (\ref{DLAx}) becomes
\begin{equation}
\left[\left(\omega+2\frac{d}{d \ln Q^2} \right) \frac{d}{d \ln Q^2}
-2C_A a_s(Q^2) A 
-\left(\omega+2\frac{d}{d \ln Q^2}\right)a_s(Q^2)\overline{P}^{(0)}(\omega)\right] D(\omega,Q^2)=0,
\label{DRAPpre}
\end{equation}
where for completeness we have also accounted for $a_s\overline{P}^{(0)}$, which can
be neglected in the following calculation of $P^{\rm DL}$.
Making the replacement in Eq.\ (\ref{replace}) in Eq.\ (\ref{DGLAPx}), taking its
Mellin transform,
$\frac{d}{d\ln Q^2}D(\omega,Q^2)=P^{\rm DL}(\omega,a_s(Q^2))D(\omega,Q^2)$,
and then substituting this into Eq.\ (\ref{DRAPpre})
gives $2(P^{\rm DL})^2+\omega P^{\rm DL}-2C_A a_s A=0$.
We choose the solution
$P^{\rm DL}(\omega,a_s)=\frac{A}{4}\left(-\omega+\sqrt{\omega^2+16C_A a_s}\right)$
since its expansion in $a_s$ yields at LO the result 
$a_s P^{{\rm DL}(0)}(\omega,a_s)=2C_A A \frac{a_s}{\omega}$,
which agrees with the LO DLs from the literature \cite{Altarelli:1977zs}. The resummed
result in $x$ space is then
$P^{\rm DL}(z,a_s)=\frac{A\sqrt{C_A a_s}}{z\ln \frac{1}{z}}
J_1\left(4\sqrt{C_A a_s}\ln \frac{1}{z}\right)$,
with $J_1$ being the Bessel function of the first kind. 

If we approximate $a_s\overline{P}^{(0)}(\omega)$ in Eq.\ (\ref{DRAPpre})
by its SLs, defined at LO to be the coefficients of $\omega^0$,
$P^{{\rm SL}(0)}_{qq}=0$, $P^{{\rm SL}(0)}_{qg}=-3C_F$, 
$P^{{\rm SL}(0)}_{gq}=\frac{2}{3}T_R n_f$ and $P^{{\rm SL}(0)}_{gg}=-\frac{11}{6}C_A-\frac {2}{3}T_R n_f$,
then if we apply the approximate result that follows from
the DLA at large $Q$,
\begin{equation}
D_{q,\overline{q}} =\frac{C_F}{C_A}D_g,
\label{DLArelforDquarkandDg}
\end{equation}
the gluon component of Eq.\ (\ref{DRAPpre}) becomes the MLLA differential 
equation \cite{Dokshitzer:1984dx}. 
Therefore we conclude that, since we do not use these two approximations,
our approach is more complete and accurate than the MLLA.

We now compare our approach to
normalized differential cross section data for light charged hadron production
from $e^+ e^- \rightarrow (\gamma,Z) \rightarrow h+X$.
We fit the gluon $D_g(x,Q_0^2)$ and the quark FFs
\begin{equation}
D_{uc}(x,Q_0^2)=\frac{1}{2}\left(D_u(x,Q_0^2)+D_c(x,Q_0^2)\right),\quad
D_{dsb}(x,Q_0^2)=\frac{1}{3}\left(D_d(x,Q_0^2)+D_s(x,Q_0^2)+D_b(x,Q_0^2)\right),
\end{equation}
where $Q_0=14$ GeV. 
Since the hadron charge is summed over, we set $D_{\overline{q}}=D_q$.
For each of these three FFs, we choose the parameterization
$D(x,Q_0^2)=N\exp(-c\ln^2 x)x^{\alpha} (1-x)^{\beta}$.
To prevent too many free parameters, we use Eq.\ (\ref{DLArelforDquarkandDg}) to fix
$c_{uc}=c_{dsb}=c_g$ and $\alpha_{uc}=\alpha_{dsb}=\alpha_g$.
Since all scales are above the bottom quark mass, we set $n_f=5$.

\begin{figure}[h!]
\includegraphics[width=8.5cm]{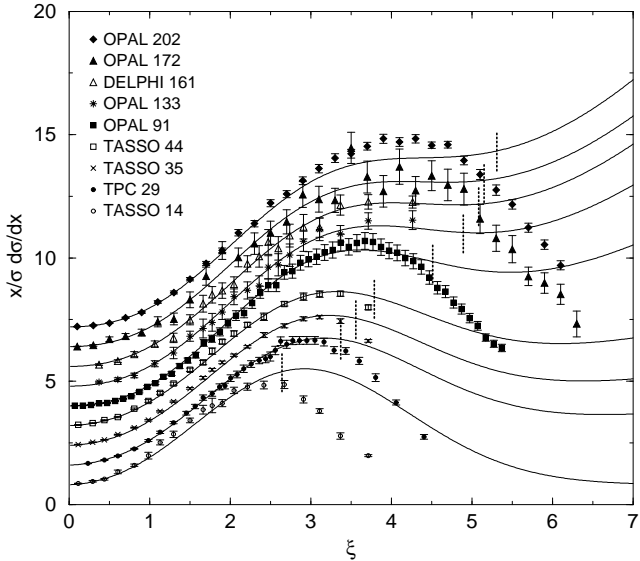}
\caption{Fit to data in the FO approach to LO.
Some of the data sets used for the fit are shown,
together with their theoretical predictions from the results of the fit. 
Only data for which 
$\xi =\ln (1/x) < \ln \sqrt{s}$ \cite{Albino:2004yg} were used, indicated by the
vertical lines. Each curve is shifted up by 0.8 for clarity.}
\label{fig1}
\end{figure}

Performing a fit using the FO approach to LO, we obtain $\chi^2_{\rm DF}=3.0$
and the results in Fig.\ \ref{fig1}. The fitted result of
$\Lambda_{\rm QCD}=388$ MeV is quite consistent with
that of other analyses, at least within the theoretical error of a factor of $O(1)$.
It is clear that FO DGLAP evolution fails in the description of the peak region and
shows a different trend outside the fit range. 

\begin{figure}[h!]
\includegraphics[width=8.5cm]{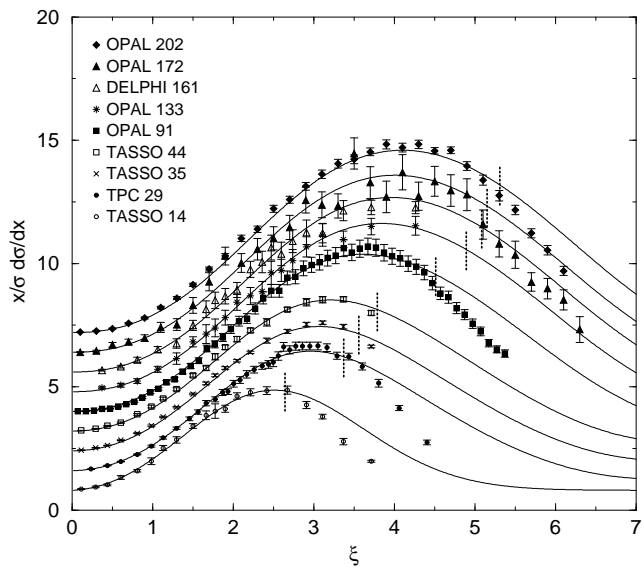}
\caption{As in Fig.\ 1, but using our approach for the evolution.}
\label{fig2}
\end{figure}

Now we perform the same fit again, but using our approach, i.e.\ Eq.\ (\ref{DGLAPx})
with the replacement in Eq.\ (\ref{replace}), for the evolution.
The results are shown Fig.\ \ref{fig2}. 
We obtain $\chi^2_{\rm DF}=2.1$, a significant improvement to the fit
above with FO DGLAP evolution. 
The data around the peak is now much better described.
The energy dependence is well reproduced up to the largest $\sqrt{s}$ value,
$\sqrt{s}=202$ GeV. We obtain a rather large $\Lambda_{\rm QCD}=801$ MeV.
We note that had we made the usual DLA (MLLA) choice
$Q=\sqrt{s}/2$ instead of our choice $Q=\sqrt{s}$ as is done in analyses
using the DGLAP equation, we would have obtained half this value for
$\Lambda_{\rm QCD}$. A treament to NLO is required to understand this problem, as
well as treatment of hadron mass effects which are important at small $x$.

In conclusion, we have proposed a single unified scheme which can describe
a larger range in $x$ than either FO DGLAP evolution or the DLA.
Our scheme allows a determination of quark and gluon FFs over a wider range of data
than previously achieved, and should
be incorporated into global fits of FFs such as that in Ref.\ \cite{Albino:2005me} 
since the current range of $0.1<x<1$ is very limited.
Our approach should be expected to improve the description of other
inclusive hadron production processes, e.g. those involving protons in the
initial state.
This work was supported in part by DFG
through Grant No.\ KN~365/3-1 and by BMBF through Grant No.\ 05~HT4GUA/4.

\end{document}